\documentclass[10pt]{emulateapj}
\usepackage{amsmath}
\usepackage{bm}
\usepackage{graphicx}

\usepackage{natbib}
\begin{document}

\title{On the Conservation of Cross Helicity and Wave Action in 
 Solar-Wind Models with Non-WKB Alfv\'en Wave Reflection}




\author{Benjamin D. G. Chandran\altaffilmark{1}, Jean
C. Perez\altaffilmark{1}, Daniel Verscharen\altaffilmark{1},
Kristopher G. Klein\altaffilmark{1}, \& Alfred Mallet\altaffilmark{1}}

\altaffiltext{1}{Space Science Center and Department of Physics,  University of New Hampshire, Durham, NH 03824;  benjamin.chandran@unh.edu}

\begin{abstract}
  The interaction between Alfv\'en-wave turbulence and the background
  solar wind affects the cross helicity
  ($\int d^3 x\, \bm{v} \cdot \bm{B} $) in two ways. Non-WKB 
  reflection converts outward-propagating Alfv\'en waves into
  inward-propagating Alfv\'en waves and vice versa, and the turbulence
  transfers momentum to the background flow.  When both
  effects are accounted for, the total cross helicity is conserved.
  In the special case that the background density and
  flow speed are independent of time, the equations of cross-helicity
  conservation and total-energy conservation can be combined to
  recover a well-known equation derived by Heinemann and Olbert that
  has been interpreted as a non-WKB generalization of wave-action
  conservation. This latter equation (in contrast to cross-helicity
  and energy conservation) does not hold
  when the background varies in time.
\end{abstract} 

\maketitle

\vspace{0.2cm} 
\section{Introduction}
\label{sec:intro}
\vspace{0.2cm} 

Approximately fifty years ago, \cite{parker65} and \cite{coleman68}
suggested that waves and turbulence play an important role in the
heating and acceleration of the solar wind.  Since that time,
observational, theoretical, and numerical studies have produced
mounting evidence that supports this suggestion. For example, in situ
measurements at heliocentric distances exceeding 0.3~AU show that
turbulent fluctuations pervade the interplanetary
medium~\citep{goldstein95a,bruno05} and that most of this turbulence
consists of fluctuations that propagate away from the Sun, consistent
with a solar origin~\citep{belcher71,tumarsch95}.  Remote observations
from the {\em Solar Optical Telescope} on the {\em Hinode} satellite
reveal the presence of Alfv\'en-wave-like motions in the low corona with
amplitudes sufficient to power the solar
wind~\citep{depontieu07}. Faraday rotation of radio transmissions from
the {\em Helios} satellite are also consistent with theoretical models
in which the solar wind is powered by an Alfv\'en-wave (AW) energy
flux~\citep{hollweg10}.

In order for AW turbulence to heat the solar wind, AW energy that is
initially in large-wavelength fluctuations must ``cascade'' to smaller
wavelengths, at which the fluctuations can efficiently dissipate. This
cascade process relies upon the interaction between
counter-propagating AWs~\citep{iroshnikov63,kraichnan65}. Because the Sun launches only
outward-propagating waves, solar-wind heating by AW turbulence
requires some source of inward-propagating AWs.

One of the most important sources of such inward-propagating waves is
non-WKB reflection~\citep{heinemann80,velli93,hollweg07}. Photospheric
motions have such long timescales that they launch AWs that have
radial wavelengths within the corona and solar wind that can be
comparable to or greater than the heliocentric distance. For such
waves, the wave phase velocity varies appreciably over one wave
length, which causes the AWs to undergo partial reflection as they
propagate away from the Sun.

A number of authors have conducted theoretical and numerical
investigations of solar-wind turbulence driven by non-WKB AW
reflection~\citep[e.g.,][]{zhou89,velli89,matthaeus99b,dmitruk02,cranmer05,verdini07,chandran09c,verdini12,perez13}. These
authors took the background solar wind to be steady, and several of
them made use of a conservation law first obtained by
\cite{heinemann80}, which \cite{heinemann80} interpreted as a non-WKB
generalization of wave-action conservation. In
this paper, we show that this conservation relation does not hold in the case of a
time-dependent background. We also show that this
conservation relation can be
obtained by combining the equations of cross-helicity conservation and
 energy conservation. To the best of our knowledge, the
equation of cross-helicity conservation has not been applied previously
to reflection-driven AW turbulence in the solar wind.
Because the equation of ``non-WKB wave-action conservation'' can be
obtained from the equation of cross-helicity conservation, and because cross
helicity is conserved regardless of whether the background varies in
time, the equation of cross-helicity conservation in some sense
generalizes the equation of ``non-WKB wave-action conservation'' to
the time-dependent regime. We note that although we allow the
flow velocity and density to vary in time, our analysis is limited to
the case in which the background magnetic field is fixed.

In Section~\ref{sec:analysis}, we describe and adopt a standard set
of approximations that has been used in previous treatments of non-WKB
reflection of Alfv\'en waves in the solar wind. For example, we assume
that the background magnetic field is quasi-radial
(Section~\ref{sec:backgroundB}) and that the fluctuations are transverse and non-compressive
(Section~\ref{sec:transverse}).  We then summarize the derivation of
the equations that describe the coupled evolution of the fluctuations and the
background flow (Section~\ref{sec:coupled}).  In
Section~\ref{sec:conservation}, we describe how the equations in
Section~\ref{sec:coupled} lead to conservation equations for
the total energy and total cross helicity. We also show
how these two conservation equations can be combined to recover the aforementioned
equation of ``non-WKB wave-action conservation'' when the background
flow is independent of time. Finally, in Section~\ref{sec:action}, we
describe how the equations in our model separately conserve the action
of inward and outward-propagating AWs in the limit of short
wavelengths and small wave amplitudes.

\section{Coupled Equations for the Fluctuating Fields and Background Flow}
\label{sec:analysis} 

We begin with the equations of ideal
magnetohydrodynamics (MHD),
\begin{equation}
\frac{\partial \rho}{\partial t} \:=\: - \nabla \cdot (\rho \bm{v}),
\label{eq:cont} 
\end{equation} 
\[
\rho\left(\frac{\partial \bm{v}}{\partial t} + \bm{v} \cdot \nabla \bm{v}\right)
\: =\:
- \nabla \left( p + \frac{B^2}{8\pi}\right) +
\frac{\bm{B}\cdot \nabla \bm{B} }{4\pi}
\]
\begin{equation} 
\hspace{1.5cm} - \frac{G M_{\sun}  \rho \bm{\hat r}}{r^2},
\label{eq:momentum} 
\end{equation} 
and
\begin{equation}
\frac{\partial \bm{B}}{\partial t} \:=\: \nabla \times \left( \bm{v} \times \bm{B}\right),
\label{eq:ind2} 
\end{equation} 
where $\rho$ is the mass density, $\bm{v}$ is the velocity,
$\bm{B}$ is the magnetic field, $p$ is the pressure, $G$ is the
gravitational constant, and $M_{\sun}$ is the mass of the Sun.
We assume that the plasma satisfies the energy equation
\begin{equation}
\frac{1}{\gamma-1}\left[\frac{\partial p}{\partial t} + \nabla \cdot
  (\bm{v} p)\right] = - p \nabla\cdot \bm{v}  - \nabla \cdot \bm{q},
\label{eq:energy} 
\end{equation} 
where $\bm{q}$ is the heat flux and $\gamma$ is the ratio of specific
heats.

\subsection{Two Models for the Background Magnetic Field}
\label{sec:backgroundB}

We set 
\begin{equation}
\bm{B} = \bm{B}_0 + \bm{\delta B},
\label{eq:Bexp} 
\end{equation} 
where the background magnetic field $\bm{B}_0$ is a fixed,
time-independent function of space.  We neglect solar rotation and
consider two models for the background magnetic field:

\vspace{0.2cm} 
\noindent {\em Model 1:} $\bm{B}_0$ consists
of exactly radial magnetic field lines filling a region that spans a
solid angle of order unity as seen from the Sun, with $B_0$ depending
only on heliocentric distance~$r$.  

\vspace{0.2cm} 
\noindent {\em Model 2:} $\bm{B}_0$ corresponds to a narrow magnetic
flux tube in which the magnetic field is nearly but not exactly
radial, which allows for super-radial divergence of the magnetic
field.  In this second case, we take the flux-tube to have an
approximately square cross section and impose periodic boundary
conditions on the edges of this cross section, as described further in
the Appendix. We also take the
opening angle~$\theta$ of the flux tube to be~$\ll 1$.

\vspace{0.2cm} 
In both models,
\begin{equation}
\bm{a} \cdot \nabla \bm{\hat{b}} = \frac{\sigma}{2}\, \bm{a}_\perp
\label{eq:sigmaid} 
\end{equation} 
for any vector~$\bm{a}$, where 
\begin{equation}
\bm{a}_\perp \equiv \bm{a} - \bm{\hat{b}} (\bm{a} \cdot \bm{\hat{b}}),
\label{eq:defaperp} 
\end{equation} 
\begin{equation}
\sigma = \nabla \cdot \bm{\hat{b}},
\label{eq:defsigma} 
\end{equation} 
and
\begin{equation}
\bm{\hat{b}} = \frac{\bm{B}_0}{B_0}.
\label{eq:defbhat} 
\end{equation} 
Equation~(\ref{eq:sigmaid}) is exact in model~1
and correct to leading order in~$\theta$ in model~2.
In both models, we are able to discard terms proportional to either $\nabla
\times \bm{\hat{b}}$ or $\bm{\hat{b}} \cdot \nabla \bm{\hat{b}}$. Such
terms vanish exactly in the case of model~1 and are much smaller
than the terms we keep in the case of model~2.

\subsection{Averages over Surfaces Perpendicular to~$\bm{B}_0$}
\label{sec:averaging0} 

For both background-magnetic-field models discussed in
Section~\ref{sec:backgroundB}, we define the ``surface
average'' of an arbitrary function~$f$, denoted~$\langle f\rangle$,
through the equation
\begin{equation}
\langle f \rangle = \frac{1}{A} \int_S dA f,
\label{eq:av} 
\end{equation} 
where $\int_S dA f$ denotes an integral of~$f$ over a surface~$S$ that
has area~$A$ and is everywhere normal to~$\bm{B}_0$. In 
model~1, $S$ is the intersection of the modeled region
with a spherical shell of some radius~$r$.
In model~2, the surface integral in Equation~(\ref{eq:av}) is confined
to the interior of the modeled magnetic flux tube. In both models, we are able to
discard terms of the form~$\langle \nabla \cdot \bm{a}_\perp \rangle$
when they arise in the derivation of the equations in
Section~\ref{sec:coupled}, where~$\bm{a}_\perp$ is everywhere
perpendicular to~$\bm{\hat{b}}$, because such terms are much smaller
than the terms we keep.  We discuss this point further in the context
of model~2 in the
Appendix. In model~1, averages of the form~$\langle \nabla \cdot
\bm{a}_\perp\rangle$ can be dropped when $\bm{a}_\perp$ contains one
or more fluctuating quantities in part because we assume that
\begin{equation}
L_{\rm c \perp} \ll r,
\label{eq:Lc} 
\end{equation} 
where $L_{\rm c \perp}$ is the correlation length of the turbulence
perpendicular to~$\bm{B}_0$.

\subsection{The Transverse, Non-Compressive Approximation}
\label{sec:transverse} 

As in Equation~(\ref{eq:Bexp}), 
we set each of $\rho$, $\bm{v}$, and $p$ equal to the sum of a
background value (denoted with a '0' subscript) and a fluctuating part
($\delta \rho$, $\delta \bm{v}$, and $\delta p$).
We define
\begin{equation}
\bm{v} = v_\parallel \bm{\hat{b}} + \bm{\delta v}_\perp,
\label{eq:defdv} 
\end{equation} 
where $\bm{\delta v}_\perp \cdot \bm{\hat{b}} = 0$.
For the scalar quantities~$v_\parallel$, $p$, and $\rho$, we define
the background quantities to be surface averages:
\begin{equation}
U \equiv v_{\parallel 0} = \langle v_\parallel \rangle \qquad 
p_0 = \langle p \rangle \qquad \rho_0 = \langle \rho \rangle.
\label{eq:averages} 
\end{equation} 
We assume that 
\begin{equation}
\nabla \cdot \bm{\delta v}_\perp = 0,
\label{eq:inc} 
\end{equation} 
\begin{equation}
\bm{\delta B} \cdot \bm{\hat{b}} = 0,
\label{eq:dBperp} 
\end{equation} 
\begin{equation}
\delta v_\parallel \equiv v_\parallel - U = 0,
\label{eq:dvpar} 
\end{equation} 
and
\begin{equation}
\delta \rho \ll \rho_0 \qquad \delta p \ll p_0.
\label{eq:NI} 
\end{equation} 
We refer to Equations~(\ref{eq:inc}) through (\ref{eq:NI}) as the
transverse, non-compressive approximation. 
Observations provide some support for this approximation.
For example, in situ measurements  show that turbulent fluctuations in
the solar wind are weakly compressive and preferentially
transverse~\citep[see, e.g.,][]{tumarsch95,goldstein95a,bruno05}.
Radio-scintillation observations further indicate that
$\delta \rho \ll \rho_0$ at heliocentric distances as small as a few
solar radii~\citep{coles89,markovskii02b,chandran09d}.  On the
other hand, transverse, non-compressive fluctuations nonlinearly
generate compressive fluctuations and longitudinal fluctuations at
some level, an effect that we neglect.
Thus, while the transverse, non-compressive approximation may apply to the bulk of the
fluctuation energy, the equations we derive in
Section~\ref{sec:coupled}  do not account for all of the physical
 processes occurring in solar-wind turbulence.

\subsection{Coupled Equations for the Fluctuating and Background
  Quantities}
\label{sec:coupled} 

In this section, we present the equations that describe the fluctuations
and background flow to leading order in~$\theta$,
$L_{\rm c \perp}/r$, $\delta \rho/\rho_0$, and $\delta p/p_0$.
To obtain an equation describing the average parallel velocity,
we take the dot product of Equation~(\ref{eq:momentum})
with~$\bm{\hat{b}}$ and then average the resulting equation over a
surface perpendicular to $\bm{\hat{b}}$ as described in
Section~\ref{sec:averaging0}. This yields
\[
\rho_0 \left(\frac{\partial U}{\partial t} + U\frac{\partial U}{\partial r} \right)
= - \frac{\partial }{\partial r} \left(p_0 + \frac{\langle |\bm{\delta B}|^2\rangle }{8\pi}\right) 
\]
\begin{equation}
+ \frac{\sigma}{2} \left(\rho_0 \langle |\bm{\delta v}_\perp|^2 \rangle - \frac{\langle | \bm{\delta B}|^2\rangle }{4\pi}\right) - \frac{G M_{\odot} \rho_0}{r^2}.
\label{eq:avpar2} 
\end{equation} 
The term proportional to~$\sigma$ on the right-hand side of
Equation~(\ref{eq:avpar2}) is the radial component of the averaged MHD Reynolds stress
$\langle \bm{\delta B}\cdot \nabla \bm{\delta B}/4\pi - \rho_0
\bm{\delta v}_\perp \cdot \nabla \bm{\delta v}_\perp \rangle$, and was
obtained previously by \cite{usmanov11,usmanov14}.

Upon multiplying Equation~(\ref{eq:avpar2})  by $\bm{\hat{b}}$ and
subtracting the resulting equation from Equation~(\ref{eq:momentum}),
we obtain the perpendicular momentum equation,
\[
\rho_0 \frac{\partial }{\partial t} \bm{\delta v}_\perp+ \rho_0 \bm{U}
\cdot \nabla \bm{\delta v}_\perp + \frac{\rho_0 \sigma
  U}{2}\bm{\delta v}_\perp = - \nabla_\perp \left(\delta \Pi\right)
\]
\begin{equation}
 +\,
\frac{\bm{B}_0 \cdot \nabla \bm{\delta B}}{4\pi}  \,+\, \frac{\sigma
  B_0 \bm{\delta B}}{8\pi} \,+\, \left(\frac{ \bm{\delta B} \cdot
  \nabla \bm{\delta B}}{4\pi} \,-\, \rho_0 \bm{\delta v}_\perp \!\cdot \nabla \delta
  \bm{v}_\perp \right)_\perp,
\label{eq:pmom1} 
\end{equation} 
where 
\begin{equation} 
\bm{U} = U \bm{\hat{b}},
\label{eq:defbmU} 
\end{equation} 
$\delta \Pi$ is the fluctuating part of $p + B^2/8\pi$, the quantity
$(\bm{a})_\perp$ is defined via Equation~(\ref{eq:defaperp}) for
arbitrary~$\bm{a}$, and $\nabla_\perp f = (\nabla f)_\perp$ for an
arbitrary scalar function~$f$.
The projection of Equation~(\ref{eq:ind2}) onto a plane perpendicular
to~$\bm{\hat{b}}$ yields 
\[
\frac{\partial }{\partial t} \bm{\delta B} = \bm{B}_0 \cdot \nabla
\bm{\delta v}_\perp -  \bm{U} \cdot \nabla \bm{\delta B} + 
\frac{\sigma}{2} ( U \bm{\delta B} - B_0 \bm{\delta v}_\perp )
\]
\begin{equation}
- \bm{\delta B} \nabla \cdot  \bm{U} + \left(
\bm{\delta B} \cdot \nabla \bm{\delta v}_\perp - \bm{\delta v}_\perp
\cdot \nabla \bm{\delta B}\right)_\perp.
\label{eq:perpind} 
\end{equation} 

Averaging Equation~(\ref{eq:cont}) as in Section~\ref{sec:averaging0}, we obtain
\begin{equation}
\frac{\partial \rho_0}{\partial t} = - \nabla \cdot (\rho_0 U
\bm{\hat{b}}).
\label{eq:cont0} 
\end{equation} 
Subtracting Equation~(\ref{eq:cont0}) from
Equation~(\ref{eq:cont}),  we find that
\begin{equation}
\frac{\partial }{\partial t}\delta \rho + \bm{\delta v}_\perp \cdot
\nabla \delta \rho = - \nabla \cdot (\delta \rho U
\bm{\hat{b}}) .
\label{eq:passive} 
\end{equation} 
Because of the transverse, non-compressive approximation
(Equations~(\ref{eq:inc}) through (\ref{eq:NI})), the density
fluctuations have no effect on the flow to leading order. 
Equation~(\ref{eq:passive}) thus
describes the evolution of passive-scalar
density fluctuations in the expanding solar wind.

We define a normalized magnetic fluctuation
\begin{equation}
\bm{\delta w} = \frac{\bm{\delta B}}{\sqrt{4\pi \rho_0}}
\label{eq:defw} 
\end{equation} 
and the Elsasser variables
\begin{equation}
\bm{z}^\pm = \bm{\delta v}_{\perp} \mp \bm{\delta w}.
\label{eq:defzpm} 
\end{equation} 
Given our sign convention in Equation~(\ref{eq:defzpm}), $\bm{z}^+$
($\bm{z}^-$) represents non-compressive, Alfv\'en-wave-like fluctuations 
that propagate in the direction of~$\bm{B}_0$ ($-\bm{B}_0$).
By combining Equations~(\ref{eq:pmom1}) and
(\ref{eq:perpind}),
we find that
\[
\frac{\partial}{\partial t}\, \bm{z}^\pm + (\bm{U} \pm \bm{v_{\rm A}})
\cdot \nabla \bm{z}^\pm =  
- \nabla_\perp (\delta \Pi) - \frac{\sigma}{2} (U \mp v_{\rm A}) \bm{z}^\mp
\]
\begin{equation}
+\frac{(\bm{z}^+ - \bm{z}^-)}{2}\left(
\nabla \cdot \bm{v_{\rm A}} \mp 
\frac{1}{2}\,\nabla \cdot \bm{U}\right)
- (\bm{z}^\mp \cdot \nabla \bm{z}^\pm)_\perp ,
\label{eq:elsasser1} 
\end{equation} 
where 
\begin{equation}
\bm{v_{\rm A}} \equiv \frac{\bm{B}_0}{\sqrt{4\pi \rho_0}}
\label{eq:defvavec} 
\end{equation} 
is the Alfv\'en velocity.
Equation~(\ref{eq:elsasser1}) was previously used
by \cite{chandran09c} and is a specialized form of the
more general Elasser-variable equation obtained  by
a number of authors \citep[e.g.,][]{zhou90,velli93,verdini07,zank12},
in which we have used Equation~(\ref{eq:sigmaid})  to
replace the quantity
$\bm{z}^\pm \cdot \nabla (-\bm{U} \pm \bm{v}_{\rm A})$ appearing in
those studies with the quantity
$(\sigma/2) \bm{z}^\pm (-U \pm v_{\rm A})$.
We rewrite Equation~(\ref{eq:elsasser1}) in terms of the Elsasser
stream functions and Elsasser vorticities in the Appendix.

\section{Conservation Laws}
\label{sec:conservation}

To obtain an equation expressing conservation of total energy, we
first take the dot product of
Equation~(\ref{eq:elsasser1})  with~$2\bm{z}^\pm$ and average over
a surface perpendicular to~$\bm{\hat{b}}$ to find
\[
\frac{\partial}{\partial t} \langle (z^\pm)^2 \rangle
+ \left(\bm{U} \pm \bm{v}_{\rm A}\right) \cdot \nabla
\langle (z^\pm)^2 \rangle
= -\sigma \left(U\mp v_{\rm A} \right)\langle \bm{z}^+ \cdot
  \bm{z}^-\rangle
\]
\begin{equation} 
\pm \langle (z^\pm)^2 - \bm{z}^+ \cdot \bm{z}^-\rangle \left(\nabla
  \cdot \bm{v}_{\rm A} \mp \frac{1}{2} \nabla \cdot \bm{U}\right).
\label{eq:fluct_energy} 
\end{equation} 
We then 
take the sum of the following equations: 
Equation~(\ref{eq:avpar2})  multiplied by~$U$;
Equation~(\ref{eq:cont0}) 
multiplied by~$U^2/2$; the ``plus version'' of
Equation~(\ref{eq:fluct_energy}) 
multiplied by $\rho_0/4$; the ``minus version'' of
Equation~(\ref{eq:fluct_energy}) 
multiplied by $\rho_0/4$;
and the average of Equation~(\ref{eq:energy}) over a surface
perpendicular to~$\bm{\hat{b}}$. This yields 
\begin{equation}
\frac{\partial {\cal E}_{\rm tot}}{\partial t} + \nabla \cdot
\bm{F}_{\rm tot} =
0,
\label{eq:econs1} 
\end{equation} 
where 
\begin{equation} 
{\cal E}_{\rm tot} = \frac{\rho_0 U^2}{2} + \frac{p_0}{\gamma-1} + \rho_0 \Phi 
+ {\cal E}_{\rm fluct}
\label{eq:edensity} 
\end{equation} 
is the surface-averaged total-energy density,
\begin{equation}
{\cal E}_{\rm fluct} = \frac{\rho_0}{4} \langle (z^+)^2 + (z^-)^2\rangle
\label{eq:Ew} 
\end{equation} 
is the energy density of the turbulent fluctuations,
\begin{equation}
\Phi = - \frac{G M_{\sun}}{r}
\label{eq:defPhi} 
\end{equation} 
is the gravitational potential, and 
\[
\bm{F}_{\rm tot} = \bm{U}\left( \frac{\rho_0 U^2}{2} + \frac{\gamma p_0
  }{\gamma -1} + \rho_0 \Phi + {\cal E}_{\rm fluct}  + \frac{\langle
    (\delta B)^2\rangle}{8\pi}\right)
\]
\begin{equation}
+ \frac{\rho_0 \bm{v}_{\rm A}}{4} \langle (z^+)^2 - (z^-)^2\rangle
+\langle \bm{q}\rangle 
\label{eq:defFtot} 
\end{equation} 
is the surface-averaged total-energy flux.

The surface-averaged cross-helicity density is
\begin{equation}
\mathcal{H}_{\rm c} = \langle \bm{v} \cdot \bm{B} \rangle = 
UB_0 + \frac{\sqrt{\pi \rho_0}}{2}\langle (z^-)^2 - (z^+)^2\rangle  .
\label{eq:defHtotal} 
\end{equation} 
We obtain the equation expressing total-cross-helicity conservation by 
adding the following equations: the ``minus version'' of
Equation~(\ref{eq:fluct_energy})  multiplied by~$\sqrt{\pi \rho_0}/2$;
the ``plus version'' of Equation~(\ref{eq:fluct_energy})
multiplied by~$ - \sqrt{\pi \rho_0}/2$; and Equation~(\ref{eq:avpar2})
multiplied by~$B_0/\rho_0$.  This yields
\begin{equation} 
\frac{\partial \mathcal{H}_{\rm c}}{\partial t} + \nabla \cdot \bm{F}_{\rm c} =
0,
\label{eq:Hccons} 
\end{equation} 
where 
\begin{equation} 
\bm{F}_{\rm c} = \bm{B}_0 \left(\frac{U^2}{2} - \frac{\langle \delta
    v_\perp^2\rangle}{2} + \Phi + h\right) + \bm{U} \langle \bm{\delta
v}_{\perp} \cdot \bm{\delta B}\rangle
\label{eq:defFc} 
\end{equation} 
is the surface-averaged cross-helicity flux, and 
\begin{equation}
h(r) = \int dr\,\frac{1}{\rho_0} \frac{dp_0}{dr} .
\label{eq:defh} 
\end{equation} 

The fact that cross helicity is conserved in the presence of non-WKB
wave reflection is perhaps surprising. Most studies of incompressible
MHD turbulence focus on the case of a stationary background, in which
the cross helicity arises entirely from the turbulent fluctuations. In
that case, converting $z^\pm$ fluctuations into $z^\mp$ fluctuations would
violate cross-helicity conservation.  In contrast, the interaction
between fluctuations and a moving, inhomogeneous, and time-dependent
solar wind changes the cross helicity via two mechanisms: non-WKB wave
reflection, which changes the cross helicity in the fluctuations, and
the transfer of momentum from the fluctuations to the background
plasma, which alters the cross-helicity content of the background
flow. The combined effect of these two mechanisms conserves the total
cross helicity in the sense of Equation~(\ref{eq:Hccons}), which, when
integrated over some arbitrary volume, implies that the change in the
total cross helicity within that volume equals the amount of cross
helicity that flows into that volume through its boundaries.

We note that total cross helicity is also conserved in weak,
homogeneous, compressible MHD turbulence, despite the fact that
interactions between Alfv\'en waves and magnetosonic waves convert
$z^\pm$ energy into $z^\mp$ energy~\citep{chandran08b}. In that
problem, there is no flow of cross helicity through the boundaries,
and the change in the cross helicity of the fluctuations is exactly
offset by the change in the cross helicity of the background. The
cross helicity of the background changes because the resonant
three-wave interactions that convert $z^\pm$ fluctuations into $z^\mp$
fluctuations simultaneously generate a small, average, background flow
parallel or anti-parallel to~$\bm{B}_0$.

We can combine the equations of cross-helicity conservation and energy
conservation by first multiplying Equation~(\ref{eq:Hccons})  by
$\rho_0 U/B_0$ and then using Equation~(\ref{eq:econs1})  to rewrite the
term
$(\rho_0 U/B) \nabla \cdot (\bm{B}_0 h) = \bm{U}\cdot \nabla p_0$ in
  terms of variables other than~$p_0$. Recalling that
$\bm{B}_0$ and $\bm{U}$ are parallel, which implies that
$\bm{B}_0 \cdot \nabla (\rho_0 U/B_0) =  - \partial \rho_0/\partial t$,
we find after some algebra that
\[
\frac{\partial}{\partial t} \left[ \frac{\rho_0 (U+ v_{\rm A})
    \langle (z^+)^2 \rangle }{4v_{\rm A}} 
- \frac{\rho_0 (U- v_{\rm A})
    \langle (z^-)^2 \rangle }{4v_{\rm A}} 
\right]
\]
\[
+\; \nabla \cdot \left[\frac{\bm{\hat{b}} \rho_0 (U + v_{\rm A})^2
    \langle (z^+)^2 \rangle}{4 v_{\rm A}}
- \frac{\bm{\hat{b}} \rho_0 (U - v_{\rm A})^2
    \langle (z^-)^2 \rangle}{4 v_{\rm A}}
\right]
\]
\begin{equation} 
= - \frac{\rho_0 \langle \bm{\delta v}_\perp \cdot \bm{\delta
    B}\rangle}{B_0}\,\frac{\partial U}{\partial t} - 
\frac{\langle (\delta v_\perp)^2\rangle}{2}\,\frac{\partial \rho_0}{\partial t}.
\label{eq:combined_cons} 
\end{equation} 
When the background plasma is steady, the right-hand side of
Equation~(\ref{eq:combined_cons}) vanishes, and
Equation~(\ref{eq:combined_cons}) is equivalent to Equation~(26) of
\cite{heinemann80}, which those authors interpreted as the
generalization of AW action conservation to the non-WKB
regime. Although \cite{heinemann80} derived their Equation~(26) for
linear waves, their Equation~(26) is also valid in the nonlinear
regime, provided
$\partial \rho_0/\partial t = \partial U/\partial t = 0$, as can be
seen from Equation~(\ref{eq:combined_cons}) above. On the other hand,
when the background plasma varies in time, the right-hand side of
Equation~(\ref{eq:combined_cons}) is in general nonzero. Thus,
Heinemann \& Olbert's~(\citeyear{heinemann80}) Equation~(26)
does not extend to the case of a time-dependent background.

\section{Wave Action}
\label{sec:action} 

\cite{bretherton68} considered the propagation of linear waves
in  slowly varying, inhomogeneous, moving media
 in the WKB limit of short wavelengths and short wave periods.
They took the waves to satisfy a dispersion relation of the form
\begin{equation}
\omega = \Omega(\bm{k}, \lambda(\bm{r},t)),
\label{eq:defomega} 
\end{equation} 
where $\lambda(\bm{r},t)$ is some slowly varying function of position and time.
The group velocity of the waves is then
\begin{equation}
\bm{c} = \nabla_k \Omega,
\label{eq:defc} 
\end{equation} 
where $\nabla_k$ denotes the gradient operator in wavenumber space,
and the frequency varies along a ray path according to the equation
\begin{equation}
\frac{d\omega}{dt} = \frac{\partial \Omega}{\partial \lambda}
\frac{\partial \lambda}{\partial t},
\label{eq:dwdt} 
\end{equation} 
where
\begin{equation}
\frac{d}{dt} = \frac{\partial }{\partial t} + \bm{c} \cdot \nabla.
\label{eq:ddt} 
\end{equation} 
\cite{bretherton68} showed that for a wide class of conservative
systems, including Alfv\'en waves in a time-dependent, inhomogeneous, moving medium,
\begin{equation}
\frac{d}{dt}\left(\frac{{\cal E}_{\rm
      w}}{\omega^\prime}\right)
+ \left(\nabla \cdot \bm{c}\right) \left(\frac{{\cal E}_{\rm
      w}}{\omega^\prime}\right) = 0,
\label{eq:actioncons} 
\end{equation} 
where $\bm{c}$ is the group velocity, 
\begin{equation} 
\omega^\prime = \omega - \bm{k} \cdot \bm{U}
\label{eq:omegaprime}
\end{equation} 
is the wave frequency measured in the local rest frame of the 
medium, $\bm{U}$ is the velocity of the medium, and ${\cal E}_{\rm w}$
is the energy density of the waves.

For the case of WKB Alfv\'en waves propagating in a radial magnetic
field, ${\cal E}_{\rm w} = \rho_0 \langle (z^\pm)^2 \rangle/4$, 
$\Omega = k_r(U \pm v_{\rm A})$, $\lambda = U \pm v_{\rm A}$,
and $\omega^\prime = \pm k_r v_{\rm A}$. Upon multiplying
Equation~(\ref{eq:actioncons})  by $\omega$ and making use of
Equation~(\ref{eq:dwdt}), we obtain
\begin{equation}
\frac{d}{dt}\left(\frac{\omega {\cal E}_{\rm w}}{\omega^\prime}\right) 
+ \left(\nabla \cdot \bm{c}\right) \left(\frac{\omega{\cal E}_{\rm
      w}}{\omega^\prime}\right) = \frac{{\cal E}_{\rm w}}{v_{\rm A}}
\frac{\partial}{\partial t} \left( v_{\rm A} \pm U\right) .
\label{eq:actioncons2} 
\end{equation} 

To see how Equation~(\ref{eq:actioncons2})  is recovered as a limiting
case in our analysis, we multiply
Equation~(\ref{eq:fluct_energy}) 
by $\rho_0(U\pm v_{\rm A})/(4v_{\rm A})$ and simplify the resulting
expression using Equation~(\ref{eq:cont0}) and the identities
$\bm{B}_0 \cdot \nabla (\rho_0 U/B_0) = -\partial \rho_0/\partial t$
and $\nabla \cdot (\rho_0 \bm{v}_{\rm A}) = - \rho_0 \nabla \cdot
\bm{v}_{\rm A} = (1/2) \bm{v}_{\rm A} \cdot \nabla \rho_0$. After some
algebra,
we obtain
\[
\frac{\partial}{\partial t} \left[ \frac{\rho_0 (U\pm v_{\rm A})
    \langle (z^\pm)^2 \rangle }{4v_{\rm A}} \right]
+ \nabla \cdot \left[\frac{\bm{\hat{b}} \rho_0 (U \pm v_{\rm A})^2
    \langle (z^\pm)^2 \rangle}{4 v_{\rm A}}\right] 
\]
\[
=\frac{\rho_0 \langle (z^\pm)^2 \rangle}{4 v_{\rm A}} \frac{\partial
}{\partial t} \left( U \pm v_{\rm A}\right) 
+ \frac{\rho_0 \langle \bm{z}^+ \cdot \bm{z}^-\rangle}{4} \times
\] 
\begin{equation} 
\left[ \left(\frac{v_{\rm A}^2 - U^2}{v_{\rm A}}\right)\left(\sigma +
      \frac{\bm{\hat{b}}\cdot \nabla \rho_0}{2\rho_0}\right) 
- \frac{(U\pm v_{\rm A})}{2\rho_0 v_{\rm A}} \frac{\partial
  \rho_0}{\partial t}\right].
\label{eq:action_cons_gen1} 
\end{equation} 
In the limit of short wavelengths and small wave amplitudes,
$\langle \bm{z}^+ \cdot \bm{z}^- \rangle \rightarrow 0$.  In this
limit, Equation~(\ref{eq:action_cons_gen1}) reduces to
Equation~(\ref{eq:actioncons2}).  When $\partial \rho_0/\partial t$,
$\partial U/\partial t,$ and $\partial v_{\rm A}/\partial t$ vanish,
subtracting the ``minus version'' of
Equation~(\ref{eq:action_cons_gen1}) from the ``plus version'' of
Equation~(\ref{eq:action_cons_gen1}) reproduces
Equation~(\ref{eq:combined_cons}) with the right-hand side of
Equation~(\ref{eq:combined_cons}) replaced by zero.

\vspace{0.2cm} 
\section{Conclusion}
\label{sec:conclusion} 

 Conservation laws play a fundamental role in the study of
  turbulence, because they are among the few analytic results that can
  be used to gain insight into the physics of turbulent systems.  For
  example, energy conservation underpins the concept of an energy
  cascade, in which nonlinear interactions among fluctuations transfer
  fluctuation energy in a loss-free manner from large scales to small
  scales.  This idea ultimately explains why the turbulent heating
  rate can be determined solely from the properties of the turbulence
  at large scales (i.e., the inertial range or the outer scale),
  regardless of the mechanisms that dissipate the energy at small
  scales.  Conservation of magnetic helicity in MHD turbulence leads
  to the concept of an inverse cascade of magnetic helicity, which
  plays an important role in turbulent
  dynamos~\citep{frisch75,pouquet76}.  In this paper, we have shown
  that a third conservation law, that of cross helicity, applies to
  non-WKB AWs and reflection-driven AW turbulence in the solar
  wind. This result is in some ways surprising, because non-WKB
  reflection converts $z^\pm$ fluctuations into $z^\mp$ fluctuations,
  thereby altering the cross-helicity content of the fluctuations. The
  total cross helicity is nevertheless conserved because the
  fluctuations exert a force on the background solar wind, which
  alters the cross-helicity content of the background flow.

  Our finding that cross helicity is conserved by non-WKB AWs and
  reflection-driven AW turbulence is important for a few
  reasons. First, it implies that cross helicity can be exchanged
  between the fluctuations and the background flow without loss.
  Second, in contrast to the equation of ``non-WKB wave-action
  conservation'' derived by \cite{heinemann80}, cross-helicity
  conservation holds even when the background flow varies in
  time. (Also, as discussed in Section~\ref{sec:conservation}, the
  equations of cross-helicity and energy conservation can be combined
  to recover Heinemann \& Olbert's~(\citeyear{heinemann80})
  conservation law when the background solar wind is
  time-independent.) Third, the coupled equations for the fluctuations
  and background flow in Section~\ref{sec:analysis} can be solved
  numerically to provide new insights into the heating and
  acceleration of the solar wind by reflection-driven AW turbulence,
  and cross-helicity conservation provides a valuable benchmarking
  tool for such simulations. There is a growing interest in numerical
  simulations of the solar wind that incorporate AW turbulence
  \citep[see, e.g.,][]{usmanov11,usmanov14,vanderholst14}, in part
  because of the upcoming launch of {\em Solar Probe Plus}. This pioneering
  mission will shed new light on the mechanisms that heat and
  accelerate the solar wind by providing the first-ever in-situ
  measurements of the solar-wind acceleration region.  By providing
  new insights into one such mechanism (reflection-driven AW
  turbulence) as well as a valuable benchmarking tool for certain
  types of numerical codes, our results may ultimately contribute to a
  deeper understanding of the solar wind's origin.  

\acknowledgements This work
was supported in part by NASA grants NNX11AJ37G and NNX15AI80G and NSF
grants AGS-1258998,  AGS-1331355, and PHY-1500041.

\appendix

\section{The Narrow-Magnetic-Flux-Tube Approximation}

As discussed in Section~\ref{sec:backgroundB}, our results apply under
either of two different assumptions about the geometry of the background
magnetic field. The background field can be either  exactly radial
throughout a region spanning a solid angle of order unity as seen from
the Sun, or the background field can be approximately radial within a
narrow magnetic flux tube centered on a radial magnetic field line.
In this appendix, we consider the case of a narrow magnetic flux tube.
We work in spherical polar coordinates and take $\theta=0$ to
correspond to the exactly radial background magnetic field line that
coincides with the axis of the magnetic flux tube. 
We restrict our analysis to a region within which
\begin{equation}
\theta  \ll 1.
\label{eq:nft} 
\end{equation} 
We assume that
\begin{equation}
  B_{0 \phi} \:=\: \frac{\partial B_{0r}}{\partial \phi} \:=\: \frac{\partial B_{0\theta}}{\partial \phi} \:=\: 0
\label{eq:Bphi} 
\end{equation} 
and define
\begin{equation}
H(r) \:=\: | \bm{B}_0(r,\theta=0)|.
\label{eq:olB} 
\end{equation} 
The condition that $\nabla \cdot \bm{B}=0$ implies that $\partial B_\theta/\partial \theta = 0$ at $\theta=0$.
We require that 
$\nabla^2 \bm{B}_0$ be finite, which implies that $\partial B_{0r}/\partial \theta  $ vanishes at $\theta = 0$,
so that 
\begin{equation}
B_{0r}(r,\theta) \:=\: H(r)\left[ 1 +  \ensuremath{\mathcal{O}}(\theta^2)\right].
\label{eq:B0r} 
\end{equation} 
The condition $\nabla \cdot \bm{B}_0 = 0$ then implies that
\begin{equation}
B_{0\theta} \:=\: - \frac{\theta}{2r} \frac{d}{dr}( r^2 H) \left[ 1 + \ensuremath{\mathcal{O}}(\theta^2)\right].
\label{eq:B0theta} 
\end{equation} 
We assume that 
\begin{equation}
\frac{r}{H}\, \frac{d H}{dr} \:\sim\: \mathcal{O}(1).
\label{eq:Blengthscale} 
\end{equation} 
Equations~(\ref{eq:Bphi}), (\ref{eq:B0r}) and (\ref{eq:B0theta}) imply that 
\begin{equation}
\bm{\hat{b}} \:=\: \left[\,\bm{\hat{r}} 
\,-\, \bm{\hat{\theta}} \,\frac{\theta}{2r H} \frac{d}{dr} (r^2 H) \,\right] \left[ 1 + \ensuremath{\mathcal{O}}(\theta^2)\right],
\label{eq:bhat} 
\end{equation} 
where $\bm{\hat{b}} \equiv \bm{B}_0/B_0$.
It follows from Equation~(\ref{eq:bhat}) that
\begin{equation}
\nabla \cdot \bm{\hat{b}} \:=\: \sigma + \mathcal{O}\left(\frac{\theta^2}{r}\right),
\label{eq:divbhat} 
\end{equation} 
where
\begin{equation}
\sigma \:=\: - \,\frac{1}{H} \,\frac{d H}{dr} .
\label{eq:sigma} 
\end{equation} 
It also follows that for any vector~$\bm{a}$, 
\begin{equation}
\bm{a} \cdot \nabla \bm{\hat{b}} \:=\: \frac{\sigma}{2} \bm{a}_\perp + \mathcal{O}\left(\frac{\theta a_\perp}{r}\right) + \mathcal{O}\left(\frac{\theta a_\parallel}{r}\right),
\label{eq:bhat_id} 
\end{equation} 
where $a_\parallel = \bm{\hat{b}} \cdot \bm{a}$ and $\bm{a}_\perp = \bm{a}
- a_\parallel \bm{\hat{b}}$.

The quantities $\nabla \times \bm{\hat{b}}$ and $\bm{\hat{b}} \cdot \nabla \bm{\hat{b}}$ relate to twist and curvature of magnetic field lines and satisfy the relations
\begin{equation}
\bm{\hat{b}} \cdot \nabla \bm{\hat{b}}=  \nabla_\perp \psi  + {\cal
  O}\left(\frac{\theta^2}{r}\right)
\label{eq:curv} 
\end{equation} 
and
\begin{equation}
\nabla \times \bm{\hat{b}} = \bm{\hat{b}} \times (\bm{\hat{b}} \cdot \nabla \bm{\hat{b}}),
\label{eq:twist} 
\end{equation} 
where 
\begin{equation}
\psi = \frac{\theta^2 r}{8} \left[ 2 \frac{d}{dr}(\sigma r) + (\sigma r -
  2)\sigma\right].
\label{eq:defpsi} 
\end{equation} 
Equation~(\ref{eq:twist}) is not a vector identity, but is exact
because of Equation~(\ref{eq:Bphi}).
It follows from Equations~(\ref{eq:Blengthscale}), (\ref{eq:divbhat}), (\ref{eq:curv}) and~(\ref{eq:twist}) that
\begin{equation}
|\nabla \times \bm{\hat{b}}| \:\sim\: |\bm{\hat{b}} \cdot \nabla \bm{\hat{b}}|
\:\sim \: \mathcal{O}\left(\frac{\theta}{r}\right) \ll |\nabla \cdot
\bm{\hat{b}}| \sim \mathcal{O}\left(\frac{1}{r}\right).
\label{eq:curl_bhat} 
\end{equation}

\begin{figure}[t]
\centerline{
\includegraphics[width=8cm]{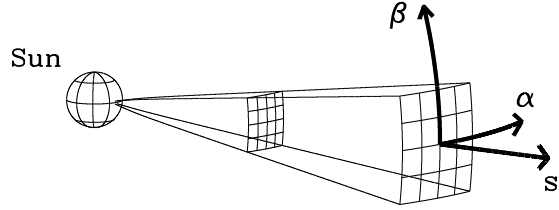}
} \caption{Clebsch coordinates $(\alpha, \beta, s)$. Surfaces of
  constant~$s$ are perpendicular to~$\bm{B}_0$. The values of $\alpha$
  and $\beta$ are constant along field lines of~$\bm{B}_0$. \label{fig:clebsch}}
\end{figure}

At several points in Sections~\ref{sec:analysis} and~\ref{sec:conservation}, we average equations over
surfaces that are everywhere perpendicular to~$\bm{B}_0$, as described
in Section~\ref{sec:averaging0}. To
specify these surfaces mathematically in the case of a narrow magnetic
flux tube in which $\bm{B}_0$ is not exactly radial, we 
introduce
the vector potential~$\bm{A}_0$ associated with the background magnetic
field
and define Clebsch
coordinates (Euler potentials) $\alpha$ and $\beta$ that are related to
$\bm{A}_0$ through the equation~$\bm{A}_0 = \alpha \nabla \beta$, which
yields
\begin{equation}
\bm{B}_0\: =\: \nabla \alpha \times \nabla \beta.
\label{eq:clebsch} 
\end{equation} 
Since $\bm{B}_0 \cdot \nabla \alpha = \bm{B}_0 \cdot \nabla \beta = 0$, $\alpha$ and $\beta$ are constant along the magnetic field lines of~$\bm{B}_0$. The particular Clebsch coordinates that we use are 
\begin{equation}
\alpha \:= \: [H(r)]^{1/2} x 
\label{eq:alpha} 
\end{equation} 
and
\begin{equation}
\beta  \:= \: [H(r)]^{1/2} y, 
\label{eq:beta} 
\end{equation} 
where $(x,y,z)$ are Cartesian coordinates, and the positive $z$ axis
coincides with $\theta=0$.  When Equations~(\ref{eq:alpha}) and
(\ref{eq:beta}) are substituted into Equation~(\ref{eq:clebsch}), the
resulting value of~$\bm{B}_0$ satisfies Equations~(\ref{eq:B0r}) and
(\ref{eq:B0theta}), as required.   We introduce a third
coordinate $s$ such that surfaces of constant~$s$ are perpendicular
to~$\bm{B}_0$, with $s = r$ at $\theta = 0$:
\begin{equation}
s \:=\: r \,-\, \frac{(x^2 + y^2)}{4 r^2 H} \frac{d}{dr} \left[ r^2 H(r)\right] + \mathcal{O}(\theta^4 r).
\label{eq:s} 
\end{equation} 
The $(\alpha, \beta, s)$ coordinate system is illustrated in 
Figure~\ref{fig:clebsch}.

For any function $f(\alpha, \beta, s)$, we define $\langle f \rangle$ to
be the average of~$f$ over a surface of constant~$s$ with $-\delta <
\alpha < \delta$ and $-\delta < \beta < \delta$, where $\delta $ is a
constant that fixes the width of the flux tube. Two
such surfaces are sketched with cross-hatched lines in
Figure~\ref{fig:clebsch}. Thus,
\begin{equation}
\langle f\rangle \:= \:
\frac{1}{A} \int_{-\delta}^{\delta} d\alpha \int_{-\delta}^{\delta} d\beta\, \frac{f}{B_0},
\label{eq:av_f} 
\end{equation} 
where
\begin{equation}
A \:=\: \int_{-\delta}^{\delta} d\alpha \int_{-\delta}^{\delta} d\beta \,\frac{1}{B_0}
\label{eq:defA} 
\end{equation} 
is the area of the averaging surface.

Equations~(\ref{eq:B0r}), (\ref{eq:B0theta}), and~(\ref{eq:s})  imply that
$|\bm{B}_0| = H(s)[1 + \mathcal{O}(\theta^2)]$. We can
thus rewrite Equation~(\ref{eq:av_f}) as
\begin{equation}
\langle f \rangle \:=\: \frac{1}{\Phi} \int_{-\delta}^{\delta} d\alpha \int_{-\delta}^{\delta} d\beta \,f  \left[1 + \mathcal{O}(\theta^2)\right],
\label{eq:avf2} 
\end{equation} 
where 
\begin{equation}
\Phi = 4 \delta^2
\label{eq:phi} 
\end{equation} 
is the magnetic flux through the averaging surface, which is
independent of~$s$. Equation~(\ref{eq:s}) implies that $|\nabla s| = 1 + \mathcal{O}(\theta^2)$, so that
\begin{equation}
\bm{\hat{b}} \cdot \nabla f  \:=\: \frac{\partial f}{\partial s} \, [1 + \mathcal{O}(\theta^2)].
\end{equation} 
It follows from Equation~(\ref{eq:avf2}) that 
\begin{equation}
\left\langle \frac{\partial f}{\partial s} \right \rangle \:=\: \frac{\partial}{\partial s} \langle f\rangle \left[1 + \mathcal{O}(\theta_0^2)\right],
\label{eq:dfds} 
\end{equation} 
where $\theta_0$ is the value of $\theta$ at the middle of one of the
edges of the averaging surface - e.g., at coordinates $(\alpha, \beta,s) = (\delta, 0, s)$.

We say that a scalar function~$g$ satisfies periodic boundary
conditions in the plane perpendicular to~$\bm{B}_0$ if it obeys the
relations
\begin{equation}
g(\delta, \beta, s)\:  = \: g(-\delta, \beta, s)
\qquad
g(\alpha, \delta,  s) \: =\:  g(\alpha, -\delta,  s)\label{eq:bc2} 
\end{equation} 
for all $\alpha \in (-\delta, \delta)$, all
$\beta \in (-\delta, \delta)$, and all~$s$. We say that a
vector~$\bm{a}$ satisfies periodic boundary conditions in the plane
perpendicular to~$\bm{B}_0$ if the vector's components
$\bm{a} \cdot \nabla \alpha$, $\bm{a}\cdot \nabla \beta$, and
$\bm{a}\cdot \nabla s$ satisfy the periodicity relations
in Equations~(\ref{eq:bc2}).  If a
vector~$\bm{a}_\perp$ is periodic in the plane perpendicular
to~$\bm{B}_0$ and satisfies~$\bm{a}_\perp \cdot \bm{B}_0 = 0$
everywhere, then Stokes' theorem can be used to show that
\begin{equation}
\langle \nabla \cdot \bm{a}_\perp \rangle\: \sim \:  \frac{\theta_0 \langle |\bm{a}_\perp|^2\rangle^{1/2}}{r},
\label{eq:avdiva} 
\end{equation} 
where we have assumed that the characteristic length scale
of~$\bm{a}_\perp$ perpendicular to~$\bm{B}_0$ is $\sim \theta_0 r$.
The rms value of $\nabla \cdot \bm{a}_\perp$ on the averaging surface
is $\sim \langle |\bm{a}_\perp|^2\rangle^{1/2} /(\theta_0
r)$.
Equation~(\ref{eq:avdiva}) thus implies that the average of
$\nabla \cdot \bm{a}_\perp$ is reduced relative to its rms value by a
factor of~$\sim \theta_0^2$.  (This reduction factor would be even
smaller if the characteristic length scale of $\bm{a}_\perp$
perpendicular to~$\bm{B}_0$ were much smaller than~$\theta_0 r$.)
This reduction enables us to drop averaged quantities of the form
$\langle \nabla \cdot \bm{a}_\perp\rangle$ in
Section~\ref{sec:analysis}, because they contribute only higher-order
corrections to the equations presented.

\section{Elsasser Stream Functions and Vorticities}

We define the 
Elsasser stream functions~$\zeta^\pm$ and the Elsasser
vorticities~$\Omega^\pm$ 
through the equations
\begin{equation}
\bm{z}^\pm = \bm{\hat{b}} \times \nabla \zeta^\pm
\qquad
\Omega^\pm = \nabla^2_\perp \zeta^\pm ,
\label{eq:defOmegapm} 
\end{equation} 
where
\begin{equation}
\nabla_\perp^2 f \equiv \nabla \cdot (\nabla_\perp f) = \nabla \cdot
\left[ \nabla f - \bm{\hat{b}} (\bm{\hat{b}} \cdot \nabla f) \right]
\label{eq:perpdiv} 
\end{equation} 
for any function~$f$. By taking the cross product of Equation~(\ref{eq:elsasser1})
with~$\bm{\hat{b}}$ and then taking the divergence of the resulting
equation, we can rewrite Equation~(\ref{eq:elsasser1}) in the form
\[
\frac{\partial}{\partial t}\, \Omega^\pm + (U \pm v_{\rm
  A})\left(\bm{\hat{b}}\cdot \nabla \Omega^\pm +
  \frac{\sigma}{2}\,\Omega^\pm\right) \:=\:
\frac{\sigma}{2}\, (-U \pm v_{\rm A}) \Omega^\mp
+ \frac{1}{2}\,\left(\nabla \cdot \bm{v_{\rm A}} \mp
\frac{1}{2}\,\nabla \cdot \bm{U}\right) (\Omega^+ - \Omega^-)
\]
\begin{equation}
- \frac{1}{2}\left(
\{ \zeta^-, \nabla^2_\perp \zeta^+\}  +
\{ \zeta^+, \nabla^2_\perp \zeta^-\}  \pm
\nabla^2_\perp \{\zeta^-, \zeta^+\}\right),
\label{eq:elsasser2} 
\end{equation}
where
\begin{equation}
\{f,g\} \equiv \bm{\hat{b}}\cdot (\nabla_\perp f \times \nabla_\perp g).
\label{eq:poisson} 
\end{equation} 
Here, we have assumed that either~$\theta \ll 1$ for the case in which
the background magnetic field corresponds to a narrow magnetic flux
tube or $L_{\rm c \perp} \ll r$ for the case in which~$\bm{B}_0$ is
radial throughout a region spanning a solid angle of order unity,
where $L_{\rm c \perp}$ is the correlation length of the fluctuations
perpendicular to~$\bm{B}_0$.  Equation~(\ref{eq:elsasser2})
generalizes Equation~(A4) of \cite{vanballegooijen11} to account for
the background flow~$\bm{U}$. Equation~(\ref{eq:elsasser2}) is the
same equation that was solved numerically by \cite{perez13}. (Note
that the minus sign on the right-hand side of their Equation~(10),
which was erroneous, was a typo in their paper, not an error in their
code.)  The form of the nonlinear term on the last line of
Equation~(\ref{eq:elsasser2}) is the same as in the RMHD equations
derived by \cite{schekochihin09}, except that our $\zeta^\pm$
corresponds to their~$\zeta^\mp$. In the homogeneous-background
limit~(in which $\sigma$, $\nabla\cdot \bm{v_{\rm A}}$, and
$\nabla \cdot \bm{U}$ vanish), Equation~(\ref{eq:elsasser2}) reduces
to Equation~(21) of \cite{schekochihin09}.

\bibliography{articles}

\end{document}